\documentclass[%
 reprint,
 amsmath,amssymb,
 aps,
pre,
]{revtex4-2}

\usepackage{graphicx}% Include figure files
\usepackage{dcolumn}% Align table columns on decimal point
\usepackage{bm}% bold math
\usepackage{hyperref}% add hypertext capabilities
\usepackage{xcolor}
\usepackage{braket}

\begin{document}

\title{A minimal quantum heat pump based on high-frequency driving and non-Markovianity}

\author{Manuel Alamo}
\email{alamo@tu-berlin.de}
\affiliation{Technische Universit{\"a}t Berlin, Institut f{\"u}r Theoretische Physik, Hardenbergstraße 36, Berlin 10623, Germany}%\date{\today}
\author{Francesco Petiziol}
\affiliation{Technische Universit{\"a}t Berlin, Institut f{\"u}r Theoretische Physik, Hardenbergstraße 36, Berlin 10623, Germany}%\date{\today}
\author{Andr\'e Eckardt}

\affiliation{Technische Universit{\"a}t Berlin, Institut f{\"u}r Theoretische Physik, Hardenbergstraße 36, Berlin 10623, Germany}%\date{\today}

\begin{abstract}
We propose a minimal setup for a quantum heat pump, consisting of two tunnel-coupled quantum dots, each hosting a single level and each being coupled to a different fermionic reservoir. The working principle relies on both non-Markovian system-bath coupling and driving induced resonant coupling. We describe the system using a reaction-coordinate mapping in combination with Floquet-Born-Markov theory and characterize its performance. 

\end{abstract}

\maketitle

\section{Introduction}

The concept of open quantum systems is based on the desire to conceptually separate a quantum system of interest from its environment, in order to find an effective description of it, without taking into account the details of the environmental degrees of freedom. 
In the ultraweak coupling limit, it is possible to derive a master equation for the dynamics of the open system of Lindblad form~\cite{Petruccione2007}, which has also been extended to periodically-driven systems~\cite{Blumel1991, Kohler1997, Grifoni1998}. This formalism entails several approximations, such as a large separation of relaxation timescales between system and bath, factorization of the system-environment quantum state, population-coherence decoupling, which are often collectively referred to as Born-Markov-secular. Large part of these assumptions are strongly challenged when the coupling with the bath is not weak anymore, in the so-called non-Markovian regime~\cite{Breuer2016, Vega2017, Talkner2020}. Non-Markovian effects have been studied, for instance, in time-delayed quantum feedback~\cite{Giovannetti1999, Debiossac2022}, dynamical decoupling sequences~\cite{Mozgunov2020, Addis2015}, photosynthetic light harvesting~\cite{Huelga2013, Chen2015} and it has been investigated, whether non-Markovianity can be exploited for improving the performance of quantum heat engines~\cite{Camati2020, Shirai2021} and the related role of memory effects~\cite{Thomas2018, Abiuso2019, Abah2020}. Non-Markovianity can be observed even in the weak coupling regime when the reservoirs are very cold, finite or structured~\cite{Sampaio2017, Camati2020}.
Several theoretical approaches have been proposed in order to treat quantum systems in these situations, beyond the ultraweak coupling approximation. Examples are time-convolutionless master equations~\cite{Petruccione2007}, hierarchical equation of motion techniques~\cite{Kato2018}, Hamiltonian of mean force~\cite{Jarzynski2017, Miller2018, Talkner2020}, pseudomode, reaction-coordinate and star-to-chain mappings~\cite{Strasberg2016, Nazir2018, Woods2014}, and more recently the generalized input-output method (GIOM) for molecular quantum junctions~\cite{Liu2020_1,Liu2020_2} and  the canonically-consistent master equation~\cite{Becker2022}, among others. 

In this paper, we consider non-Markovianity not as an undesired reality one has to deal with, but rather as an asset that is harnessed to build a quantum heat pump. We will not only use non-Markovianity to improve the performance of the device, but make it an indispensable part of its working principle. This allows for a minimal design of the heat pump, based on two quantum states only, by using bath degrees of freedom as a resource. 
Starting from previous theoretical proposals~\cite{Rey2007, Strasberg2016, Roy2020} and recent experimental realizations~\cite{Brantut2013, Krinner2015, Krinner2017} the proposed heat pump is
realized with two coupled single-level quantum dots, each strongly coupled to a large fermionic reservoir. We show that this two-level configuration, assisted by a time-periodic drive, can work as a heat pump that relies on non-Markovian effects induced by the strong coupling with the environment. Namely, the working principle of the heat pump will actively involve bath degrees of freedoms, as they are described within the formalism of reaction coordinate (RC) mapping~\cite{Strasberg2016, Nazir2018, Strasberg2018, Restrepo2019, Sztrikacs2021}, which are vital for establishing a finite heat current from the colder to the warmer reservoir. Via the RC mapping, the system is extended to an effective four-level system comprising the two original quantum-dot levels and one pseudomode per reservoir, with the pseudomodes, in turn, weakly coupled to residual baths. In a suitable parameter regime, the four-level system can be arranged in a level structure forming two independent transport channels coupled to the residual reservoirs. In order to control the transport of low and high energy particles through the system, the intra-dot coupling is then modulated periodically in time, realizing drive-assisted tunnelling through the channels. While the weak coupling to the residual baths justifies a treatment according to Floquet-Born-Markov-secular theory~\cite{Blumel1991, Kohler1997, Grifoni1998, Roy2020} for the whole four-level extended system, the reaction-coordinate mapping allows us to still capture strong-coupling non-Markovian effects, which are crucial for the functionality of the pump.
Our proposal aims to contribute to the theoretical discussion and provides new alternatives to the already existing one-dimensional quantum heterostructure that have been experimentally implemented in recent years. 

Minimal quantum heat engines and pumps in the non-Markovian regime have also been investigated before for driven two-level systems coupled to bosonic reservoirs~\cite{Kurizki2004, Kolar2012, Kurizki2013, Mukherjee2020, Xu2022}. In particular, the role of non-Markovianity to enhance the performance of heat pumps has been recently proposed based on the anti-Zeno effects~\cite{Mukherjee2020, Xu2022}. Our work treats fermionic reservoirs and exploits non-Markovianity arising from a fixed strong coupling to the baths, rather than from a time-dependent system-bath coupling, by including reservoir's degrees of freedom into the system through the reaction-coordinate mapping. This enables the realization of a two-way transport channel that carries heat with zero net particle transport (requiring at least four energy levels), thus leaving the mean reservoir occupation unchanged. The use of fermionic baths makes our proposal closer to recent experiments, e.g., in cold-atom systems, where fermionic reservoirs of cold atoms connected by a controllable channel have been realized~\cite{Brantut2012, Brantut2013, Krinner2015}, or quantum dots in solid-state platforms~\cite{Blanter2000, Lu2003, Milburn2001, Rey2007}. Fermionic baths probed around the Fermi energy feature a much less dramatic variation of the thermal occupation as compared to bosonic ones, for which this variation is exasperated by bosonic enhancement at lower energies. A fermion bath then gives a more uniform response to the system, namely it induces quantum-jump rates in the system of comparable strength. This, in turn, may favour the stability of the regime where the system works as a heat pump, requiring less assumptions on the bath spectral density and on the operating regime of the pump. The methodology pursued here to obtain a heat pump beyond the ultraweak system-bath coupling, building on the RC  mapping, may be extended to the case of bosonic baths, where, at the technical level, a different mapping is needed, requiring a different Bogoliubov transformation on the bath degrees of freedom to single out the RC mode and yielding a different spectral density of the residual bath~\cite{Nazir2018}.

The working principle underpinning the heat pump proposed here is recapitulated in Sec.~\ref{sec:working_principle}. The construction of the necessary level scheme and transport channels is then discussed in Sec.~\ref{sec:config}, where the reaction-coordinate mapping and the Floquet approach are detailed. The performance of the heat pump is then analyzed in Section~\ref{sec:performance}, also in the presence of drive imperfections. Thermal transport is achieved and the time evolution of the bath's temperature is studied revealing cooling process via energy pumping on short time-scales. Conclusions are then drawn in Sec.~\ref{sec:conclusion}.

\section{Working Principle} \label{sec:working_principle}
The principle of the heat pump is transporting high energy fermions from a cold to a warm reservoir and transporting low energy fermions in the opposite direction, resulting in a net heat transport from the cold reservoir to the warm one. The design of the heat pump involves two independent transport channels, as proposed in Refs.~\cite{Roy2020, Kohler2004}, which we distinguish as \emph{upper} ($+$) and  \emph{lower} ($-$). Each channel transfers particles in both ways between both reservoirs, connecting states of different energy via driving (or ``photon’’) assisted tunneling. This results in a net heat transport while maintaining zero net charge current. 

\begin{figure}
\includegraphics[width=\linewidth]{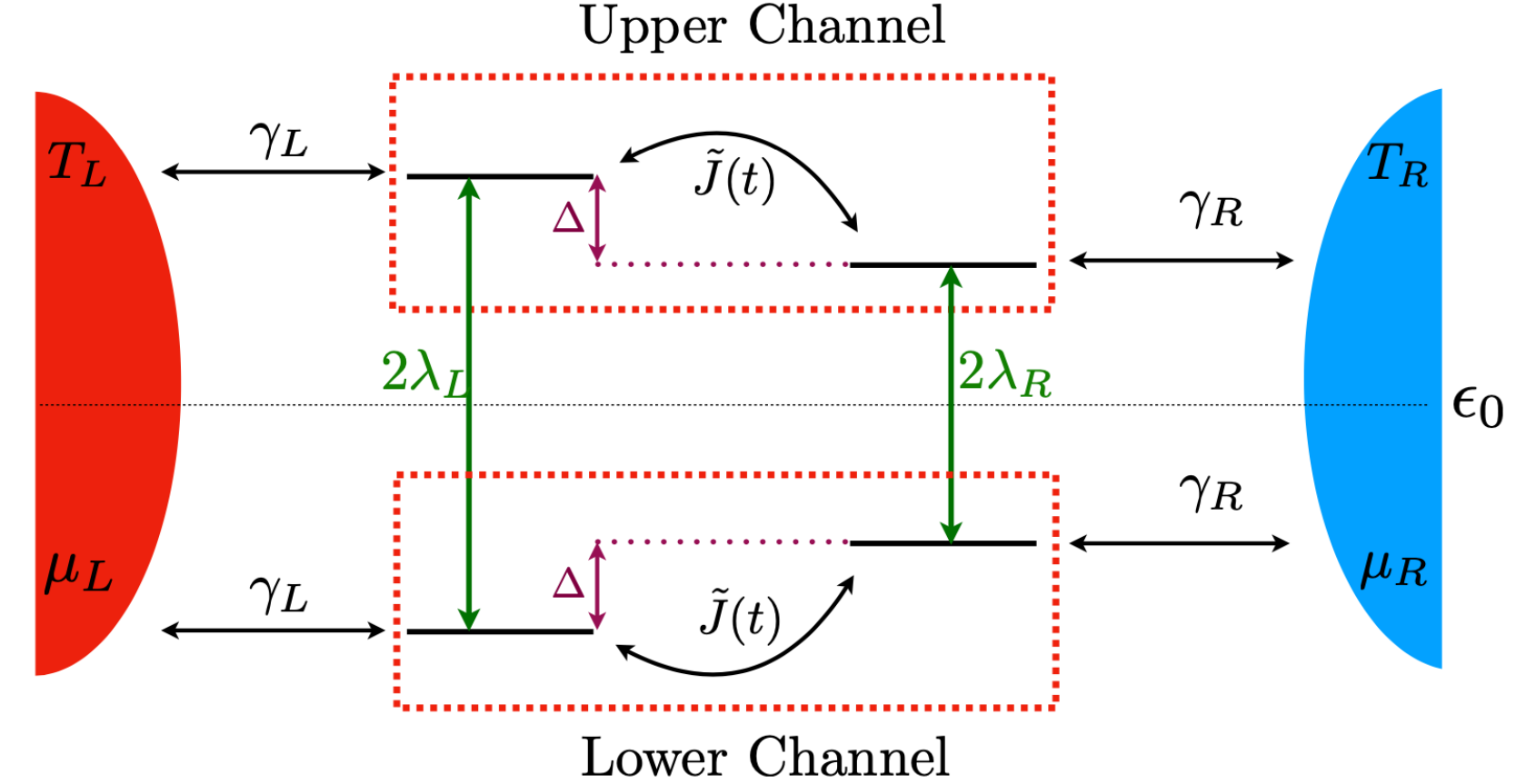}
\caption{Two-channel quantum heat pump configuration. In this situation, electrons in the upper channel can tunnel from the cold to the hot reservoir via absorbing energy from the drive, whereas fermions at lower energies can be trasported from the hot to the cold reservoir via the emission of energy quanta into the drive.}
\label{setup}
\end{figure}

Each channel is composed of two tunnel-coupled single-particle energy levels, one of which is coupled to the hot left ($L$), while the other coupled to the cold right ($R$) reservoir. The energy configuration of the four levels is shown in Fig.~\ref{setup}: both the two levels coupled to the $L$ reservoir ($E_{\pm,L}$) and the two levels coupled to the $R$ reservoir ($E_{\pm,R}$) are symmetrically detuned from a reference energy $\epsilon_0$ (see Fig.\ref{setup}),
\begin{equation}
E_{\pm, R} = \epsilon_0\pm \lambda_R\hspace{0.5cm}\text{and}\hspace{0.5cm} E_{\pm, L} = \epsilon_0\pm(\lambda_R + \Delta).
\end{equation}
The energy offset $\Delta = \lambda_L-\lambda_R$ between two channel states is much larger than the tunnelling strength $J_0$, such that tunneling across the channels is effectively suppressed. In order to induce energy-selective transport through the device, tunneling is then assisted by driving resonantly with a time-periodic force. We consider in particular a periodic modulation of the interchannel coupling of the form $J_1\cos(\omega t)$, where $J_1<J_0$ is the driving amplitude and $\omega$ corresponds to the driving angular frequency. Moreover, we choose the driving energy to be resonant with the energy offset in the channel, i.e., $m\hbar\omega =\Delta$ for some integer $m$. In such a situation, fermions of energy $\epsilon_0 + \lambda_R$ can absorb $m$ energy quanta $\hbar\omega$ when passing from the cold to the hot reservoir via the upper channel, whereas fermions at lower energies $\epsilon_0-\lambda_R$ can emit energy $m\hbar\omega$ into the drive when moving from the cold to the hot reservoir via the lower channel. The converse energy exchange with the drive occurs for the hot-to-cold transport.

A steady-state situation with vanishing net particle and energy flow between the reservoirs is achieved when the Fermi-Dirac distribution function $f_{T,\mu}(\epsilon) = 1/[e^{(\epsilon-\mu)/T}+1]$ of the right reservoir at energy $\epsilon_0 \pm \lambda_R$ equals the distribution function of the left reservoir at energy $\epsilon_0 \pm (\lambda_R + \Delta)$, providing an heuristic estimate of the regime where the system is expected to work as a heat pump~\cite{Roy2020}. Namely, assuming an equal chemical potential $\mu_L=\mu_R=\epsilon_0$ for the two baths, heat is transported from the cold to the hot reservoir until no net flow is reached for a temperature ratio
\begin{equation}
\frac{T_L}{T_R} = \frac{\lambda_R + \Delta}{ \lambda_R}  \label{temp_equation}.
\end{equation} 
Deviations from this ideal scenario arise due to driving-assisted processes also between system and bath, where energy quanta (``photons’’) $\hbar\omega$ are absorbed or emitted by the drive~\cite{Roy2020}.

In the present paper, we will show how to build the heat pump described in this Section, using \emph{two} system levels only. In order to achieve this, we harness the non-Markovian coupling to the bath, which allows us to use also reservoir degrees of freedom as a resource for realizing the necessary transport channels. This can be understood within the formalism of reaction-coordinate mapping explained in the following. 

\begin{figure*}
\includegraphics[width=\textwidth]{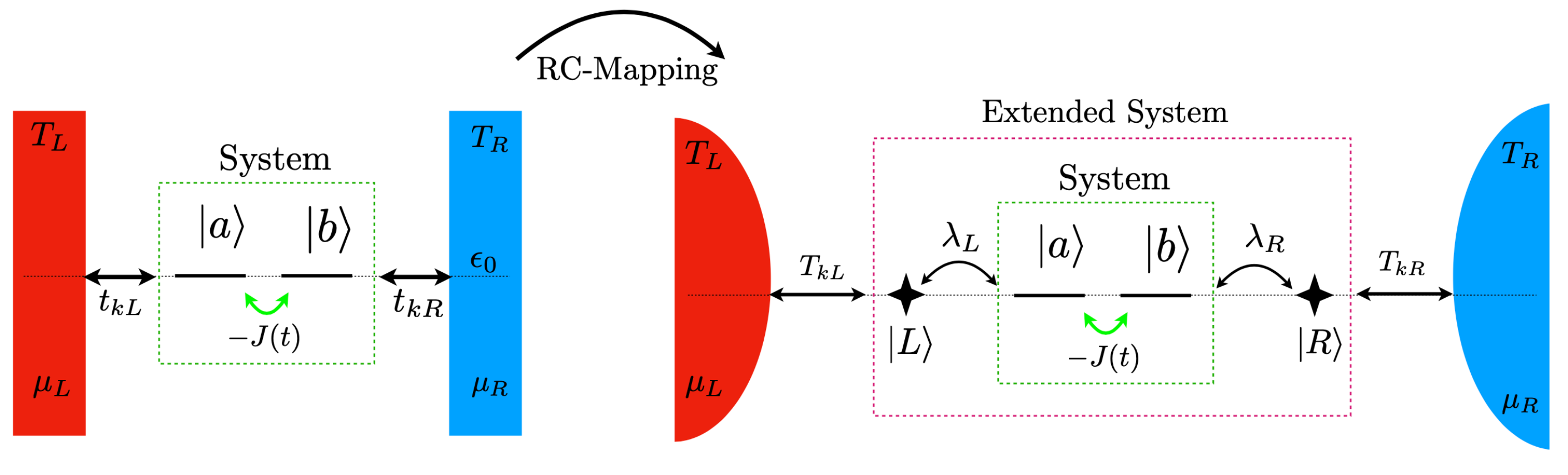}% Here is how to import EPS art
\caption{Reaction Coordinate mapping of the original system. The original two single level Quantum Dots coupled to thermal baths at different couplings is enlarged to a four single level system coupled to residual reservoir characterized by a residual  spectral density $\tilde{\mathcal{J}}(\omega)$. If as a result of the mapping, the residual baths are described by a Markovian spectral density and are weakly coupled to the extended system, then the Born-Markov approximation remains valid and the formalism of the Markovian master equation can be applied.}\label{RC Mapping}
\end{figure*}
 
\section{Heat-pump configuration} \label{sec:config}

\subsection{Extended system via reaction-coordinate mapping}
In order to effectively engineer the ideal level structure discussed in Sec.~\ref{sec:working_principle}, we consider a system composed of two single-level Quantum Dots (QD) (see the left panel of Fig.~(\ref{RC Mapping}), described by the periodically-driven time-dependent Hamiltonian
\begin{equation}
 \hat{H}_S(t) = \epsilon_a \hat{d}^\dagger_a\hat{d}_a + \epsilon_b \hat{d}^\dagger_b\hat{d}_b -J(t) \left(\hat{d}^\dagger_a\hat{d}_b + \mathrm{H.c}\right).
\end{equation}
The operators $\hat{d}_\alpha^\dagger$ and $\hat{d}_\alpha$ ($\alpha=a, b$) describe the creation and annihilation of a fermion in the left or right level, $\epsilon_\alpha$ is the on-site energy of the QD and $J(t)$ is a time-periodic tunneling matrix element between the dots,
\begin{equation} \label{driving}
J(t) = J_0 + J_1\cos(\omega t).
\end{equation}
In order to study two-terminal transport, each QD is coupled to a left$(L)$ and right$(R)$ thermal fermionic bath, characterized by a temperature $T_\nu$, a chemical potential $\mu_\nu$, and a spectral density
\begin{equation}
\mathcal{J}_\nu(\omega) = 2\pi\sum_{k}|t_{k,\nu}|^2\delta(\omega-\epsilon_{k,\nu}) \label{SD},
\end{equation} 
 with $\nu = L,R$. The baths and their coupling to the QDs are described by the Hamiltonians $\hat{H}^{(\nu)}_B$ and $\hat{H}^{(\nu)}_I $, respectively, which read
\begin{eqnarray}
\hat{H}^{(\nu)}_B &=& \sum_{\nu, k}\epsilon_{k,\nu}\hat{c}^\dagger_{k,\nu}\hat{c}_{k,\nu}, \\
\hat{H}^{(\nu)}_I &=& \sum_{k,\nu}\left(t_{k,\nu}\hat{d}_{\nu}\hat{c}^\dagger_{k,\nu} + t^*_{k,\nu}\hat{d}^\dagger_{\nu}\hat{c}_{k,\nu} \right). \label{eq:HIqd}
\end{eqnarray}
The operators $\hat{c}_{\nu, k}$, $\hat{c}^\dagger_{\nu,k}$ represent the annihilation and creation of fermions in the baths, $\epsilon_{k,\nu}$ the energy of the bath modes and $t_{k,\nu}$ describes the tunneling amplitudes between the QDs and the baths in Eq.~\eqref{SD}. The full system-bath Hamiltonian is then given by
\begin{equation}
\hat{H}(t) = \hat{H}_S(t) + \sum_{\nu = L,R}\left[\hat{H}^{(\nu)}_I+ \hat{H}^{(\nu)}_B \right].\label{Hamiltonian}
\end{equation}

The system here described may be implemented experimentally in cold-atoms setups as realized in Refs.~\cite{Brantut2012, Brantut2013, Krinner2015, Hausler2017}. For instance, in~\cite{Brantut2012, Brantut2013} a fermionic cold-atom cloud is divided into two macroscopic reservoirs separated by a channel through a repulsive laser beam focused at the center of the cloud, which confines few atoms to a quasi-2D channel. The specific potential landscape considered here, realizing the two-site channel and including the time-periodic modulation, may be imprinted using established techniques to engineer spatio-temporally structured light-shift potentials by means of, e.g., digital mirror devices~\cite{Ott2016}.

The first step for creating a two-channel device from the two-level structure and the baths is to enlarge the system's size by including two more single-site energy levels. Instead of modifying the system itself, we propose to exploit the bath degrees of freedom for this purpose. This can be achieved by considering a rather strong system-bath coupling in combination with baths of rather narrow spectral width. Using the so-called reaction coordinate (RC) mapping~\cite{Strasberg2016,Nazir2018, Strasberg2018, Restrepo2019}, one can show that under these assumptions the system is effectively coupled to one effective level (pseudomode) per bath, given by a superposition of bath modes. The RC is weakly coupled to the remaining bath degrees of freedom, which can then be treated using Floquet-Born-Markov theory. 

The RC Mapping consists in linearly transforming the fermionic bath operators $\hat{c}_{k}$ into new fermionic operators $\hat{C}_{k}$ by applying a unitary transformation $M$, $MM^\dagger=1$,
\begin{equation}
\hat{C}_{k,\nu} = \sum_{r}M_{kr}\hat{c}_{r,\nu}.
\end{equation}
One particular mode is defined by the system-bath coupling and called the reaction coordinate. It is given by imposing the following condition on the first row of $M$,
\begin{equation}
\lambda^*_{\nu}\hat{C}_{1,\nu} = \sum_k t^*_{k,\nu}\hat{c}_{k,\nu},
\end{equation}
with $\lambda_{\nu}^*$ a constant parameter such that $\lbrace \hat{C}_{1,\nu}, \hat{C}_{1,\nu}^\dagger\rbrace = 1$, which then reads~\cite{Strasberg2018, Nazir2018, Restrepo2019}
\begin{equation}
|\lambda_\nu|^2 = \frac{1}{2\pi}\int\,d\omega\mathcal{J}_\nu(\omega)\label{lambda1}.
\end{equation}
As a consequence, the Hamiltonian of the extended system (including the original system and the RCs) is given by
\begin{multline} 
\hat{\tilde{H}}_S(t) = \hat{H}_S(t) + \tilde{\epsilon}_L\hat{C}^\dagger_L\hat{C}_L + \tilde{\epsilon}_R\hat{C}^\dagger_R\hat{C}_R\\
+ \lambda_L(\hat{d}^\dagger_L\hat{C}_L + \hat{C}^\dagger_L\hat{d}_L ) +  \lambda_R(\hat{d}^\dagger_R\hat{C}_R + \hat{C}^\dagger_R\hat{d}_R ) 
\label{eq:HS_transf}
\end{multline}
with the RC on-site energy 
\begin{equation}
\tilde{\epsilon}_\nu =\frac{1}{2\pi\lambda^2_\nu}\int\,d\omega \omega \mathcal{J}_\nu(\omega).
\end{equation}
In this way the original system is coupled only to the RC, which, in turn, couples to the remaining part of both baths. These residual baths are described by  the Hamiltonian 
\begin{equation} \label{eq:HB_transf}
\hat{\tilde{H}}_B =  \sum_{k,\nu}E_{k,\nu}\hat{C}^\dagger_{k,\nu}\hat{C}_{k,\nu},
\end{equation}
where the energies $E_{k,\nu}$ are obtained imposing $l\neq1$ and $m\neq1$ that $\sum_k\epsilon_{k\nu}M_{lk}M^*_{rk} = \delta_{lr}E_{l\nu}$~\cite{Strasberg2018}.
Both residual baths interact with the extended system via the interaction Hamiltonian
\begin{equation} \label{eq:HI_transf}
\hat{\tilde{H}}_I = \sum_{k,\nu}\left(T_{k,\nu}\hat{C}_\nu\hat{C}^\dagger_{k,\nu} + T^*_{k,\nu}\hat{C}_{k,\nu}\hat{C}^\dagger_\nu\right),
\end{equation}
where the couplings are given by $T_{k\nu} = \sum_r \epsilon_{r\nu} t_{r\nu} M_{kr}/\lambda_\nu$.

We remark here that the fact that the system now couples to the reaction coordinate only [Eq.~\eqref{eq:HS_transf}], and not directly to the residual bath, is a key feature obtained on purpose through the construction of a suitable RC mapping. This does not mean that the system only couples to one `physical' bath mode, namely the RC is a coherent superposition of all the bath modes the system is coupled to. The RC is coupled to the system and the remaining bath degrees of freedom, which, for the residual bath, consist of superposition modes that are chosen to be orthogonal to both the RC and to each other and which are not coupled to the system. 
%it couples to all physical modes [see Eq.~\eqref{eq:HIqd}], but the unitary (more generally, the Bogoliubov) transformation in the bath underpinning the RC mapping is designed, such that the coupling to all modes maps to a coupling to a single mode, in turn coupled to a residual bath. 
Although not needed in our analysis, this procedure can in principle be iterated, converting the coupling of all modes to the system into a chain where the system couples to one mode only, which in turn couples to another mode only, and so forth---a so-called star-to-chain mapping~\cite{Woods2014,Nazir2018}.

So far, we have just rewritten the total Hamiltonian, without making any approximation. However, for strong system-bath coupling, previously it would have not been justified to  deriving a master equation for the system by employing Floquet-Born-Markov theory. However, for the enlarged system described by Eq.(\ref{eq:HS_transf}), given by the original system and the RC for the system alone, the coupling to the residual bath $T_{k,\nu}$ can now be small, so that we can employ Floquet-Born-Markov theory on this level. This is the case for a sufficiently narrow spectral density $\mathcal{J}_\nu(\omega)$, as will be shown below. 

For the residual baths, we obtain the spectral density
\begin{equation}
\tilde{\mathcal{J}}_\nu(\omega) = 2\pi\sum_k|
T_{k,\nu}|^2\delta(\omega-E_{k,\nu}).
\end{equation}
The system-bath Hamiltonian is then overall transformed into
\begin{equation}
\hat{\tilde{H}}(t) = \hat{\tilde{H}}_S(t) +\sum_{\nu = L,R}\left[ \hat{\tilde{H}}^{(\nu)}_I + \hat{\tilde{H}}^{(\nu)}_B\right],
\end{equation}
with $\hat{\tilde{H}}_S(t)$, $ \hat{\tilde{H}}^{(\nu)}_B$ and $ \hat{\tilde{H}}^{(\nu)}_I$ given in Eqs.~\eqref{eq:HS_transf}, \eqref{eq:HB_transf}, and \eqref{eq:HI_transf}, respectively.
It describes four effective QDs: the original two QDs in the center coupled by a time-dependent matrix element $-J(t)$, each of them coupled to a new degree of freedom (RC) via the time-independent coupling $\lambda_\nu$ (see Fig.~\ref{RC Mapping}).

Following Eq.~\eqref{lambda1}, it is clear that a concrete parametrization of $\mathcal{J}_\nu(\omega)$ is needed. In order to include non-Markovian effects into the model and to have a narrow bath we have chosen a structured-shape Lorentzian spectral density of the form~\cite{Nazir2018}
\begin{equation}
\mathcal{J}_\nu(\omega) = \frac{\Gamma_\nu\eta^2_\nu}{(\omega-\epsilon_\nu)^2+\eta^2_\nu}\label{Lorentzian},
\end{equation}
centered around the QD-energy $\epsilon_\nu$ with width $\eta_\nu$ and effective coupling strength $\Gamma_\nu$.  In the following we will choose $\epsilon_L = \epsilon_R = \epsilon_0$, such that the QDs are in a symmetric configuration and both spectral densities are centered around the same reference energy. The parameter $\eta_\nu$ controls the characterization of the baths, resulting in non-Markovian baths descriptions for finites values of $\eta_\nu$, whereas the Markovian limit is obtained in the limit $\eta_\nu\to\infty$~\cite{Nazir2018, Restrepo2019}.
Using Eq.(\ref{Lorentzian}) the residual bath parameters are given by
\begin{eqnarray}
|\lambda_\nu|^2 = \frac{\Gamma_\nu\eta_\nu}{2}, \qquad 
\tilde{\epsilon_\nu}=  \epsilon_\nu, \qquad \tilde{\mathcal{J}}_\nu = 2\eta_\nu \equiv \gamma_\nu\label{lambda},
\end{eqnarray}
with $\tilde{\mathcal{J}}_\nu$ a constant Markovian spectral density, and where we have defined the new system-bath coupling strength $\gamma_\nu$. 
Note, thus, that the RC mapping allowed us to convert the non-Markovian two-level problem into a Markovian one for the extended system: We started with reservoirs described by a structured spectral density with a defined and sharp peak around a specific frequency in Eq.~\ref{Lorentzian} and after performing the RC Mapping we ended up with a flat spectral density $\tilde{\mathcal{J}}_\nu$ that admits a fast decay in the corresponding correlation functions characteristic of a Markovian bath. The coupling to the residual bath is proportional to the bandwidth of the original bath. If it is sufficiently narrow, the residual bath can be treated using a weak-coupling Markovian master equation. This statement and the results presented in the following do not rely on the Lorentzian spectral density. We assume a Lorentzian bath only for simplicity, since it can be treated analytically and gives rise to a constant effective spectral density. 
In the following we will thus assume that the coupling strength between the RC and the residual baths is weak such that it allows us to define and implement a Floquet-Markovian master equation formalism ~\cite{Blumel1991, Kohler1997, Grifoni1998, Hone2009, Vorberg2015, Petruccione2007}. 
 
\subsection{Transport channels}
Once the four-level system has been obtained, the second step in the construction of the two independent transport channels is to realize the symmetrical level configuration described in Sec.~\ref{sec:working_principle}. For this purpose, we consider the situation, in which the coupling $\lambda_\nu$ to the RCs is strong as compared to the tunnelling couplings $J_0$ and $J_1$ in the QD, $J_0, J_1\ll\lambda_\nu$, and we consider the simple case $\epsilon_\alpha = \tilde{\epsilon_\nu} = \epsilon_0$. In this regime, we treat $J_i$ as a perturbation and diagonalize the unperturbed part of the Hamiltonian, obtaining unperturbed single-particle eigenstates that are symmetric and antisymmetric superposition of the original QD $\ket{a}$, $\ket{b}$ and RC $\ket{L}$, $\ket{R}$ states, respectively. They read
\begin{equation}
|L_\pm \rangle= \frac{|L\rangle \pm |a\rangle}{\sqrt{2}},\qquad
|R_\pm \rangle=  \frac{|R\rangle \pm |b\rangle}{\sqrt{2}}.\label{unpert_states}
\end{equation}
The corresponding unperturbed left and right energies are given by $E_{\pm, L} =\epsilon_0 \pm\lambda_L$ and $E_{\pm, R} =\epsilon_0 \pm\lambda_R$ respectively, where the positive (negative) energies correspond to the upper (lower) levels in the configuration. In this new basis, the level configuration is given by two `left' levels separated by an energy off-set of $2\lambda_L$ and two `right' levels separated by an energy difference of $2\lambda_R$. In the single-particle basis $\{\ket{L_+}, \ket{L_-}, \ket{R_+},\ket{R_-} \}$, the system Hamiltonian now reads
\begin{eqnarray} \label{eq:Hprime}
\hat{H}_S(t) &=& \epsilon_0\mathbb{I}\nonumber\\
&+& \frac{1}{2}\begin{bmatrix}
2\lambda_L & 0 & -J(t) & -J(t) \\
0 & -2\lambda_L & -J(t) & -J(t) \\
- J(t) & -J(t) &2\lambda_R & 0 \\
-J(t) & -J(t) & 0 & -2\lambda_R
\end{bmatrix},
\end{eqnarray}
with $J(t)$ modulated as in Eq.~\eqref{driving} and $\mathbb{I}$ the identity matrix. By choosing the driving energy $\hbar\omega$ to be resonant with the gap $\Delta=\lambda_L-\lambda_R$, and assuming that off-resonant couplings can be neglected, the target configuration of Sec.~\ref{sec:working_principle} and Fig.~\ref{setup} is attained. In the following (Sec.~\ref{sec:performance}), we will also analyse the case of slightly detuned driving, namely $\Delta = \hbar\omega + \delta$, and its impact of the performance of the heat pump. Note that, as compared to the configuration studied in Ref.~\cite{Roy2020}, the periodic modulation addresses tunnelling couplings among the unperturbed energy levels, rather than the unperturbed energies themselves. This equally achieves the goal of activating energy-selective tunnelling, but it produces a different micromotion dynamics not strongly dressing also the QD-RC coupling. 

Following the reasoning of Sec.~\ref{sec:working_principle}, the driven extended system is expected to work as an energy-selective channel: fermions can transit between the two baths via the two channels while increasing or decreasing their energy by absorbing or emitting, respectively, energy quanta $\hbar\omega$ from the drive. In particular, high-energy fermions in the hot left reservoir can emit $\hbar \omega$ and tunnel to the cold right reservoir via the upper channel, while the opposite process involves an absorption of $\hbar \omega$. In the lower channel, low-energy fermions from the left reservoir absorb $\hbar \omega$ to transit to the right reservoir. According to Eq.~\eqref{temp_equation} and setting $\mu_l = \mu_R$, the mentioned steady state situation is reached when~\cite{Roy2020}
\begin{equation}
\frac{T_L}{T_R} = \frac{\lambda_L}{\lambda_R} >1.\label{performance_rate}
\end{equation}
We quantitatively confirm this intuitive statement in the following.

\subsection{Floquet States} \label{sec:floquet}

The periodic time dependence of $\hat{H}_S(t)$ of Eq.~\eqref{eq:Hprime} is treated using the Floquet formalism~\cite{Sambe1973, Anisimovas2015}. Single-particle Floquet quasienergies $q_\alpha$ and Floquet modes $|u_\alpha(t)\rangle$ are computed numerically by diagonalizing the one-period time evolution operator. We here also derive analytical approximations in the limit of large driving frequency $\omega$. The effective high-frequency Hamiltonian can be determined by first performing a gauge transformation to a ``rotating"  frame of reference,
\begin{equation}
\hat{H}_S(t) \to \hat{V}^\dagger(t)\hat{H}_S(t)\hat{V}(t)-i\hbar\hat{V}^\dagger(t)\frac{d}{dt}\hat{V}(t), \nonumber
\end{equation}
 with time-periodic unitary operator $V(t) = \exp\left[- i\omega t (\ket{L_+}\!\bra{L_+} - \ket{L_-}\!\bra{L_-})\right]$,
resulting in the rotating Hamiltonian
\begin{eqnarray}
\hat{H}'_{\mathrm{eff}} &=& \epsilon_0\mathbb{I}\nonumber\\
&+&\frac{1}{2}\begin{bmatrix}
\lambda_R & 0 & -\tilde{J}(t) & \tilde{J}(t) \\
0 &-\lambda_R & \tilde{J}^*(t)& -\tilde{J}^*(t) \\
-\tilde{J}^*(t)& \tilde{J}(t) &\lambda_R & 0 \\
\tilde{J}^*(t) & -\tilde{J}(t) & 0 &-\lambda_R
\end{bmatrix},
\end{eqnarray}
with $\tilde{J}(t) \equiv e^{-i\omega t}J(t)$. After performing the transformation, all matrix elements are small compared to $\hbar \omega$, so that the state of the system hardly evolves during each driving period. This brings us into the position to perform a rotating-wave approximation by averaging over the Hamiltonian. We arrive at the effective time independent Hamiltonian
\begin{eqnarray}
\hat{H}_{\mathrm{eff}} &=& \epsilon_0\mathbb{I}\nonumber\\
&+&\frac{1}{2}\begin{bmatrix}
\lambda_R & 0 & -J_1/2 & 0 \\
0 & -\lambda_R & 0 & -J_1/2 \\
-J_1/2 & 0 &\lambda_R & 0 \\
0 & -J_1/2 & 0 &-\lambda_R
\end{bmatrix},
\end{eqnarray}
where tunnelling between states $|L_\pm\rangle$ to $|R_\mp\rangle$ is off-resonant and therefore not present in the effective time-independent Hamiltonian.
The corresponding Floquet modes associated with the upper (`$\uparrow$') and lower (`$\downarrow$') channel can thus be approximated with
 \begin{subequations} \label{eq:floquet_modes}
 \begin{eqnarray} 
 \ket{\uparrow_{\pm}(t)} & =  \frac{1}{\sqrt{2}} (e^{-i\omega t}\ket{L_+} \mp \ket{R_+}), \\
 \ket{\downarrow_{\pm}(t)} & = \frac{1}{\sqrt{2}} (e^{i\omega t}  \ket{L_-} \mp \ket{R_-}),
 \end{eqnarray}
 \end{subequations}
 and they have quasienergies
 \begin{equation}
 q_{\uparrow\pm} =(\epsilon_0 + \lambda_{R}) \pm J_1/4, \qquad q_{\downarrow\pm} = (\epsilon_0-\lambda_{R}) \pm J_1/4.
 \end{equation}
 The quasienergies are thus symmetrically arranged around an average energy of $ \epsilon_0+\lambda_R$ for the upper channel, and of $\epsilon_0-\lambda_R$ for the lower channel. In the following, we will label Floquet quasienergies and modes with a single index $\alpha$ such that $\{q_\alpha\}_{\alpha=1,\ldots,4} = \{q_{\uparrow+}, q_{\uparrow-}, q_{\downarrow+}, q_{\downarrow-}\}$ and $\{\ket{u_\alpha}\}_{\alpha=1,\ldots,4} = \{\ket{\uparrow_+}, \ket{\uparrow_-}, \ket{\downarrow_+}, \ket{\downarrow_-}\}$.

\subsection{Master Equation}

In order to obtain a master equation in Lindblad form we use the Floquet formalism of the previous Sec.~\ref{sec:floquet} together with the secular approximation~\cite{Petruccione2007}. 
The weak coupling regime considered here, where the coupling between the extended system and the residual baths $\gamma_\nu$ represents the smallest energy scale in the system, $\gamma_\nu \ll J_0,J_1\ll \Delta$, justifies a treatment of the residual bath within a Floquet-Born-Markov-Secular framework~\cite{Blumel1991, Kohler1997, Grifoni1998}. Under these circumstances, the open dynamics is described by a Markovian many-body master equation~\cite{Roy2020, Vorberg2015}. The system's steady state is given by a time periodic density matrix  $\hat{\rho}(t) = \sum_\alpha p_\alpha|n_\alpha(t)\rangle\langle n_\alpha(t)|$ which is diagonal with respect to the time-periodic Floquet-Fock states $|n_\alpha(t)\rangle$ which describe the occupation of the $\alpha$th single-particle Floquet mode $\ket{u_\alpha(t)}$.

The time-independent probabilities $p_\alpha$, corresponding to the diagonal elements of $\hat{\rho}(t)$, are determined by golden-rule-like contributions of birth ($*$) and death ($\dagger$) processes from the left and right residual reservoirs, in which the system exchanges $m$ energy quanta $\hbar\omega$ with the drive, i.e.,
\begin{equation}
R^\eta_{\alpha} = R^\eta_{\alpha, L} + R^\eta_{\alpha, R}\hspace{0.5cm}\text{and}\hspace{0.5cm} R^\eta_{\alpha, \nu} = \sum_{m=-\infty}^{+\infty}  R^{\eta (m)}_{\alpha, \nu} , 
\end{equation}
with $\eta = *, \dagger$ and $\nu = L,R$. Explicitly, the rates read
\begin{eqnarray}
R^{\eta (m)}_{\alpha, \nu} &=& \frac{2\pi}{\hbar}\left|{C}_{\nu,\alpha}^{(m)}\right|^2\nonumber\\
&\quad&\cdot D_\nu(q_\alpha + m\hbar\omega)f^\eta_{\nu, T_\nu, \mu_\nu}(q_\alpha + m\hbar\omega),
\end{eqnarray}
where
\begin{equation} \label{eq:mat_el_m}
{C}_{\nu,\alpha}^{(m)} = \frac{1}{T}\int^T_0 \,dt e^{im\omega t} \langle 0|\gamma_\nu\hat{C}_\nu|u_\alpha(t)\rangle,
\end{equation}
where $q_\alpha$ is the quasienergy of the $\alpha$th Floquet mode, and where we have considered the fermionic distribution function $f^*_{T,\mu}(\epsilon) = \lbrace\exp([\epsilon-\mu]/T)+1\rbrace^{-1}$ and $f^\dagger_{T,\mu}(\epsilon) = 1-f^*_{T,\mu}(\epsilon) $ and the bath's density of states $D_\nu(\epsilon) =\rho_\nu\Theta(\epsilon)$. Here, $\rho_\nu$ is a constant value above a minimum energy $\epsilon$ that characterize the bath and $\Theta$ is the Heaviside step function. 

In the steady state, the mean occupations of the Floquet modes obey
\begin{equation}
\frac{d}{dt}\langle \hat{n}_\alpha\rangle  = R^{*}_{\alpha}(1-\langle \hat{n}_\alpha\rangle) - R^\dagger_{\alpha}\langle \hat{n}_\alpha\rangle = 0,
\end{equation}
so that $\langle \hat{n}_\alpha\rangle = (1+R^\dagger_{\alpha}/R^*_{\alpha})^{-1}$. Using this solution we can obtain the particle and heat currents as~\cite{Roy2020}
\begin{eqnarray} \label{currents}
\dot{N}_\nu &=&\sum_\alpha\left[R^\dagger_{\alpha, \nu}\langle \hat{n}_\alpha\rangle - R^*_{\alpha, \nu}(1-\langle \hat{n}_\alpha\rangle)\right]\nonumber\\
\dot{E}_\nu &=& \sum_{\alpha, m}\left[R^{\dagger(m)}_{\alpha, \nu}\langle \hat{n}_\alpha\rangle - R^{*(m)}_{\alpha, \nu}(1-\langle \hat{n}_\alpha\rangle)\right]\nonumber\\
& &\quad \times (q_\alpha + m\hbar\omega).
\end{eqnarray}
The matrix elements~\eqref{eq:mat_el_m} can be estimated using the analytically approximated Floquet modes of Eqs.~\eqref{eq:floquet_modes}. We find
\begin{eqnarray}
C_{L, \uparrow\pm}^{(m)} =\gamma_L \delta_{m,1}/2, \quad C_{L \downarrow \pm}^{(m)} = -\gamma_L\delta_{m,-1}/2, \\
C_{R, \uparrow \pm}^{(m)} =\mp \gamma_R \delta_{m,0}/2, \quad C_{R \downarrow, \pm}^{(m)} = \pm\gamma_R \delta_{m,0}/2,
\end{eqnarray}
which are valid in the resonant and high-$\omega$ limit.

\begin{figure}
\includegraphics[width=9cm]{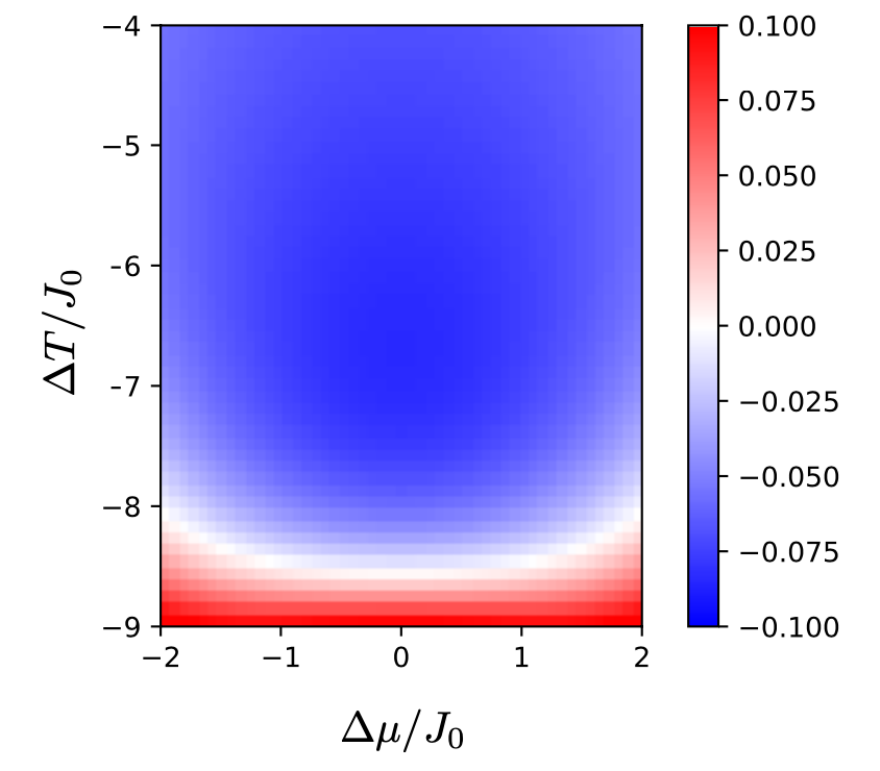}% Here is how to import EPS art
\caption{Heat Pump performance captured by the change rate of the right temperature $\dot{T}_R$ as function of  $\Delta T = T_R-T_L$ and $\Delta \mu = \mu_R-\mu_L $for resonant driving frequency $\hbar\omega/J_0 = (\lambda_L-\lambda_R)/J_0 \sim 17$. The parameter offset is $ J_1 = 0.7J_0$,  $ \Gamma_L = 10000J_0$, $ \Gamma_R = 200J_0$, $ \eta_\nu = 0.08J_0$, $ \rho_L = 10\rho_R$. The temperature and the chemical potential of the left bath are kept constant with $\mu_L = 50\epsilon_0$ and $T_L = 0.2\mu_L$.}\label{Heat_pump}
\end{figure}

\section{Heat pump performance} \label{sec:performance}

The performance and efficiency of the heat pump is characterized by analyzing the rate $\dot T_R$ at which the right bath’s temperature changes when the difference in temperature $\Delta T = T_R - T_L$ and chemical potential $\Delta \mu = \mu_R-\mu_L$ is changed.  This information is obtained by inverting the matrix relation
\begin{equation}
\begin{pmatrix}
\dot{N}_\nu\\
\dot{E}_\nu
\end{pmatrix}
 =
 \begin{pmatrix}
 \partial_{\mu_{\nu}}N_\nu & \partial_{T_{\nu}}N_\nu\\
 \partial_{\mu_{\nu}}E_\nu & \partial_{T_{\nu}}E_\nu
 \end{pmatrix}
 \begin{pmatrix}
 \dot{\mu_\nu}\\
 \dot{T}_\nu
 \end{pmatrix},\label{matrix}
\end{equation}
where ${\dot N}_\nu$ and ${\dot E}_\nu$ are given in Eq.~\eqref{currents}, and the matrix is computed from the expressions
\begin{eqnarray}
N_\nu(T_\nu, \mu_\nu) &=& \int^\infty_0\,d\varepsilon\rho_\nu f_{T_\nu, \mu_\nu}(\varepsilon),\nonumber\\
E_\nu(T_\nu, \mu_\nu) &=& \int^\infty_0\,d\varepsilon\varepsilon\rho_\nu f_{T_\nu, \mu_\nu}(\varepsilon).
\end{eqnarray}
In the following, we will use the tunnelling parameter $J_0$ as the unit of energy and time is measured in units of $\tau = \hbar/\gamma_R^2\rho_R$. If the couplings to the residual baths are identical, $\gamma_R = \gamma_L$ [$\gamma_\nu$ is defined in Eq.~\eqref{lambda}], then the dynamics depends only on the ratio $\alpha \equiv \rho_L/\rho_R$ representing the reservoir sizes.

Following Eq.~\eqref{performance_rate}, the configuration is roughly expected to  extract heat from the cold bath in the regime 
\begin{equation} \label{eq:estimate}
0>\Delta T>T_L\left(\frac{\lambda_R}{\lambda_L} -1\right).
\end{equation}
The time variation of the right-bath's temperature is reported in Fig.~\ref{Heat_pump}. We can observe an extended region, where even for negative $\Delta T$, heat is transported from the cooler to the hotter bath (blue area) for driving amplitudes $\hbar\omega/J_1\sim 24$, thus showing a small deviation from the expected theoretical predictions $\Delta T/J_0 \sim -8.58 $ obtained from Eq.~\eqref{eq:estimate}. 
Deviations of this approximated picture can be attributed to ``photon"-assisted tunnelling processes where energy quanta $m\hbar\omega$  with $|m|>1$ are absorbed or emitted by the drive, when tunneling from the extended system into the residual bath, producing driving-assisted heating associated with the micromotion present in Floquet quantum systems.

The advantages of engineering the heat pump using bath degrees of freedom are at this point evident: the performance of the heat pump depends on the energy ratio between the levels in the channels [see Eq.~\eqref{performance_rate}], which for our parametrization of the spectral density, ultimately depends explicitly on the interaction of the original QD and the original reservoirs $\Gamma_\nu$. According to Eq.~\eqref{lambda}, an increase in $\Gamma_\nu$ only increases the coupling $\lambda_\nu$ between the QD and the reaction coordinate, leaving unaffected the coupling between the reaction coordinate and the residual bath $\gamma_\nu$. This increase then still allows for the implementation of a Floquet-Born-Markov-Secular master equation formalism to study the full open quantum system, while capturing non-Markovian effects on the original single-level quantum dot system. This flexibility allows us to find parameters that reach a heat pump regime away from the ultra-weak coupling limit between the original system-bath configuration (but regarding the coupling between extended system and residual bath, we are always in the ultra weak coupling regime). 

\begin{figure}
\includegraphics[width=\linewidth]{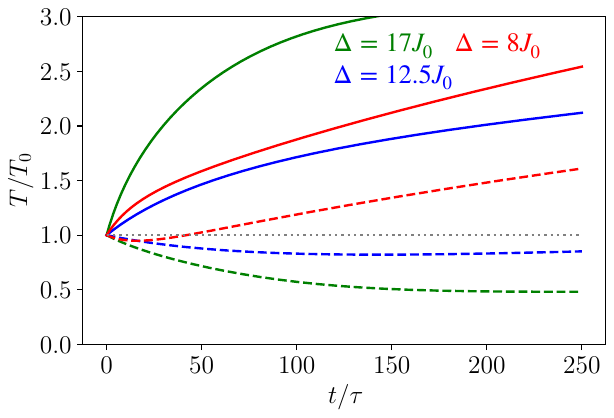}
\caption{Time evolution of the temperatures of both baths for different intra-channel energy difference $\Delta$, with $\lambda_L$ fixed at $\lambda_L=20$. We observe better efficiencies at short times when $\Delta$ is sufficiently large. Dashed lines correspond to the right bath and solid lines to the left bath.  We have considered the initial conditions: $\epsilon_0 = 50, \mu_L = \mu_R = \epsilon_0$, $T_L = T_R = T_0 = 0.2\mu_0$, $\rho_L/\rho_R = 1$ and driving amplitude $J_1 = 0.7J_0$.}\label{Temp_Delta}
\end{figure}

The heating behavior is shown in Fig.~\ref{Temp_Delta} where the time evolution of the temperature of the bath is studied by inverting the matrix in Eq.~\eqref{matrix} and then integrating over time. The solid lines represent the hot left reservoir and the dotted lines the cold right one (the parameters are given in the caption).
From Fig.~\ref{Temp_Delta} and Eq.~\eqref{performance_rate} we observe that the cooling regime requires two conditions: Large coupling strengths $\Gamma_\nu\gg J_0$ between the original system-bath configuration and that these couplings must be different, such that $\Delta = \lambda_L-\lambda_R >1$. 
The efficiency of the heat pump in the cooling regime improves when  $\Delta$ increases, entering the regime of large $\Gamma_\nu$ and moving away from the weak-coupling regime between the original system-bath configuration, highlighting the key role of the RC mapping in the performance of the heat pump. The temperature at which the heat pump operates steadily is also reached more quickly as $\Delta$ increases.

\begin{figure}
{\includegraphics[width=\linewidth]{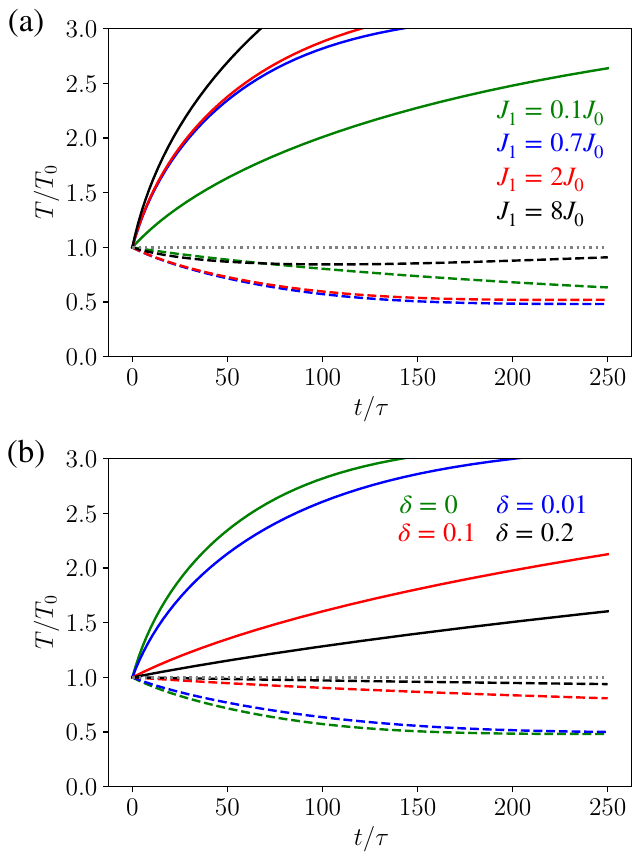}}
 \caption{a) Time evolution of the temperatures for different driving strength and $\delta = 0$. Dashed lines correspond to the right bath and solid lines to the left bath. We see that the best performance is obtained for weak driving strength and for sufficient large driving amplitude the device stops working efficiently as a heat pump. b) Time evolution of the temperatures for different values of the detuning $\delta$ and $J_1 = 0.7J_0$. We see that finite values of $\delta$ alter the efficiency of the heat pump and for sufficiently large detuning the heat pump becomes inefficient.
 We have considered the initial conditions: $\epsilon_0 = 50, \mu_L = \mu_R = \epsilon_0$, $T_L = T_R = T_0 = 0.2\epsilon_0$, $\rho_L/\rho_R = 1$ and $\Delta = 17J_0$ in both figures.}
 \label{Temp_det}
\end{figure}

Similar behaviour for the efficiency is observed as a function of the driving amplitude. In Figure~\ref{Temp_det}(a), the variation of the bath temperatures for different driving strengths $J_1$ is shown. The cooling regime is obtained for small amplitudes and reaches its minimum temperature of $T_R\sim T_0/2$ for $J_1 = 0.7J_0$ and remains stable within a finite window until the driving amplitude becomes strong enough for the device to stop working as a heat pump. 
We can associate the break down of the operation after long times to driving-assisted processes, where particles tunnel from system to bath via $m$-`photon' processes involving multiple energy quanta $m\hbar \omega$ from the drive. Such effects are beyond the rotating wave approximation for the effective Hamiltonian. 

To further benchmark the robustness of the driving-assisted heat transport in the presence of drive imperfections, in Fig.~\ref{Temp_det}(b) we analyze the impact of a detuning $\delta$ on the driving frequency that makes the desired photon-assisted tunnelling slightly off-resonant, namely such that
\begin{equation}
\hbar\omega = \Delta(1-\delta).
\end{equation}
For finite small detuning $\delta$ and small enough driven amplitudes $J_1<J_0$, the system still works as a heat pump decreasing the temperature of the right bath to a minimum of $T_{R_\text{min}}/T_0 \sim 0.5$ for $\delta = 0.01$. Transport within the channels is still present under detuned driving but with a minimal temperature larger than the one obtained in resonance, as one may expect. When increasing $\delta$ significantly, the off-resonant modulation reduces the efficiency of the energy transfer between the system and the drive, decreasing the rate of energy absorption or emission and therefore breaking down the working principle of the quantum heat pump. This can be observed when $\delta$  is increased to a value of $\delta=0.2$. For sufficiently large $\delta$ at weak driving amplitudes the drive is not able to activate the tunnelling anymore, which thus remains effectively switched off. As a consequence, the heat pump stops working completely, heating both reservoirs even for small values of the driving strength.

\section{Conclusion} \label{sec:conclusion}
We have shown that a system composed of two single-level quantum dots can work as a heat pump, when it is strongly coupled to two different non-Markovian baths, while a periodic modulation enables energy-selected particle transport. By effectively enlarging the system size exploiting bath degrees of freedom using the Reaction-Coordinate Mapping, we were able to construct two independent energy-selective photon-assisted transport channels. In this setup, high-energy fermions from a cold bath can be transported to a hot bath by absorbing $m$ energy quanta $\hbar\omega$ from the drive, whereas low energy fermions emit $m$ energy quanta $\hbar\omega$ into the drive when moving from the cold to the hot reservoir using the lower channel (see Fig.~\ref{setup}). Under certain assumptions about the spectral density of the original bath, we reach a scenario where the extended system is weakly coupled to the residual Markovian reservoir but the original two-level system remains strongly coupled to the selected reservoir degree of freedom, allowing the implementation of a Floquet-Born-Markov-Secular formalism. 

The heat pump regime is obtained away from the ultra-weak coupling between the original system-bath configuration, indicating that the heat pump is made possible by the non-Markovian character of the baths. This was verified, for example, in Fig.~\ref{Temp_Delta}, where the best performance is reached for larger system-reaction coordinate coupling, attaining a minimal temperature of $T_{R_{\text{min}}}/T_0\sim0.38$ for resonant driving amplitude $J_1 = 0.5J_0$. The efficiency of the heat pump was also studied in the presence of a finite detuning on the drive, showing a robustness of the working mechanism under driving imperfections. Indeed, a moderately large detuning still enables heat transport, though with an increase of the minimal attainable temperature.
The proposed device represents an interesting playground where non-Markovian and strong coupling effects can be studied by using recent experimental techniques for microstructuring quantum scenarios. As a future direction, this device can be also extended to the study of interacting systems, where several proposals of many-body themal machines have been addressed in recent years~\cite{Herrera2017,Niedenzu2018,Skelt2019,Mukherjee2021}
.
\begin{acknowledgments}
We acknowledge insightful discussions with Gernot Schaller and Dario Poletti. This research was funded by the German Research Foundation (DFG) via the Research Unit FOR 2414, under Project No. 277974659.
\end{acknowledgments}

\bibliography{apssamp} 
%\bibliography{biblio}  

\end{document}